\documentclass[11pt]{article}

\usepackage[T1]{fontenc}
\usepackage[utf8]{inputenc}
\usepackage{lmodern}
\usepackage[margin=1in]{geometry}
\usepackage{amsmath,amssymb,amsthm}
\usepackage{graphicx}
\usepackage{booktabs}
\usepackage{array}
\usepackage{xcolor}
\usepackage{tikz}
\usetikzlibrary{arrows.meta,positioning}
\usepackage[numbers,sort&compress]{natbib}
\usepackage[colorlinks=true,linkcolor=blue!50!black,citecolor=blue!50!black,urlcolor=blue!50!black]{hyperref}

\graphicspath{{figures/}}
\newcommand{\plainlink}[2]{{\hypersetup{urlcolor=.}\href{#1}{#2}}}

\theoremstyle{plain}
\newtheorem{proposition}{Proposition}
\theoremstyle{definition}
\newtheorem{definition}{Definition}
\theoremstyle{remark}

\newcommand{\E}{\mathbb{E}}
\newcommand{\Prob}{\mathbb{P}}
\newcommand{\Var}{\operatorname{Var}}
\newcommand{\indic}{\mathbf{1}}
\newcommand{\cS}{\mathcal{S}}
\newcommand{\cQ}{\mathcal{Q}}

\title{Measuring AI Search Visibility for AEO and GEO:\\
What Citation Metrics from ChatGPT,\\
Google AI Overviews and Perplexity Measure}

\author{%
\plainlink{https://www.elementera.com/team/soroush-mehri?r=iTZzRC}{Soroush Mehri} \qquad \plainlink{https://www.elementera.com/team/dr-mehrdad-mehrbod?r=iTZzRC}{Mehrdad Mehrbod} \qquad \plainlink{https://www.elementera.com/team/mojtaba-ebadi?r=iTZzRC}{Mojtaba Ebadi} \qquad \plainlink{https://www.elementera.com/team/dr-maziar-gomrokchi?r=iTZzRC}{Maziar Gomrokchi}\\[4pt]
\plainlink{https://www.elementera.com/?r=iTZzRC}{Elementera AI}\\[4pt]
\small Equal contribution.\\
\small Correspondence: Soroush Mehri, \texttt{soroush.mehri@elementera.com}\\
\small Project page: \plainlink{https://www.elementera.com/research?r=iTZzRC}{\texttt{www.elementera.com/research}}}

\date{Preprint, November 2026}

\begin{document}
\maketitle

\begin{abstract}
AI search visibility is now tracked by counting which sources AI answer engines such as ChatGPT, Google AI Overviews and Perplexity cite, often for answer engine optimization (AEO) and generative engine optimization (GEO). This paper explains what those citation numbers measure and when two of them can be compared. Every such number, whether a citation rate, a citation stability figure or a brand's AI visibility score, is an average of one underlying quantity: the probability that a source is shown for a query, through a given capture path, on a given occasion and draw. Two such numbers are rarely comparable, because each averages over different draws, occasions, queries, capture paths and engine sub-queries. We give a measurement model that separates a source missing from one draw from a source leaving the set the engine uses, define what citation stability and the admission stage of the citation funnel should mean, and prove what a study design can identify. Separating random noise from real change requires at least two draws of each query per occasion. Normalizing by that noise floor leaves a free-intercept decay fit unchanged. Intervals that resample only queries leave out between-occasion variance, so they cover the true value less often as more queries are added. We name five layers of averaging, including the engine's expansion of a user query into sub-queries, which no outside observer can condition on. The citation list shown with an answer is shaped by both the sources the answer was generated from and other features of the answer, so it is not a noisy copy of those sources, and conditioning on it can bias estimates in ways that more data and better parsing do not remove. A simulation study illustrates each result. We report which source fields four public AI answer surfaces expose and give a reporting checklist. We claim nothing about the effect of AEO or GEO actions.
\end{abstract}

\section{Introduction}
\label{sec:intro}

AI answer engines now sit between many searchers and the web. Google's AI Overviews and AI Mode, ChatGPT with web search, Gemini with search grounding and Perplexity each return a generated answer together with a list of cited sources. Publishers, marketers and researchers measure those lists to track AI search visibility, often under the names answer engine optimization (AEO) and generative engine optimization (GEO). The measurements take many forms: the share of queries on which a source is cited, the overlap between the sources cited on two days, the share of cited sources that do not rank in the organic top ten, or a single visibility score per brand. Studies of these quantities report figures that differ widely, and the differences are usually read as facts about the engines.

\paragraph{In brief.} A citation number from an AI answer engine is an average, and two such numbers can disagree only because they average over different things. Separating random noise from real change in which sources an engine cites requires at least two draws of the same query on the same occasion. Intervals that resample only queries become too narrow when results on the same occasion move together. On the public surfaces we captured, the source list shown with an answer is written with the answer, so it does not show which sources the engine admitted and did not cite. No outside observer can condition on how an engine expands a query into sub-queries, so every reported citation number includes that hidden average.

This paper starts from a different reading. A citation observation is a draw from a stochastic process, taken through a particular capture path on a particular occasion, for a particular query, at a particular grain (a URL, a host, or a registrable domain). The quantity that a repeated-query design can estimate is a conditional probability for one such cell. Every number that is reported about ``a source'', ``an engine'' or ``citation stability'' is obtained from cell-level quantities by averaging over some of these coordinates, by collapsing several draws or occasions into one observation, and by choosing a population of queries. Two reported numbers can therefore disagree because they average over different things, with no disagreement about the engines at all.

We do not claim that our own numbers escape this. They cannot, for a reason that is specific to answer engines. The query in our cell is the query a user types. Google documents that AI Overviews and AI Mode ``may use a `query fan-out' technique'', which it describes as ``issuing multiple related searches across subtopics and data sources'' \citep{googleaifeatures}. When an engine expands a user query into sub-queries, the probability that a source is rendered for the user query is itself an average over an engine-internal, unobserved and time-varying distribution of sub-queries. Conditioning on the user query does not end the chain of marginalizations; the engine performs one more on the observer's behalf. Our position is therefore not that we condition where others marginalize. Every reported citation number is a marginal. The requirement we argue for is that each marginalization be named, including the ones performed by the engine, and that a study state which of them its design can and cannot remove.

Two further problems follow from the structure of the observable. First, an absence is ambiguous: a source that is not rendered on a draw may have left the set of sources the engine uses for the query, or may still be in that set and not have been drawn this time. No statement about how long a citation persists is well posed until these two causes are separated, and a design with one draw per occasion cannot separate them. Second, studies of the citation funnel factor the probability of being cited into a probability of being retrieved or admitted and a probability of being cited given admission. That factorization assumes that every cited source was admitted, and it needs an observation of admission. On the public surfaces we have captured, the fields that look like a retrieval pool behave as attribution lists written with the answer. We argue that the rendered list is a common effect of admission and of answer-level factors. Conditioning on it is conditioning on a collider \citep{greenland2003collider,hernan2004selection,elwert2014endogenous}.

\paragraph{Contributions.}
\begin{enumerate}
\item A measurement model for repeated observation of answer engines that separates per-draw inclusion from latent membership, and an estimand family over the rendered surface: a same-occasion noise floor, across-occasion retention under a declared union convention, a stable core reported as a threshold family, and a recurrence response (Sections~\ref{sec:setting} and~\ref{sec:model}).
\item Identification results by design (Section~\ref{sec:identification}): draw noise and fast latent churn are observationally equivalent with one draw per occasion (Proposition~\ref{prop:k}); noise-floor normalization does not change a free-intercept decay fit (Proposition~\ref{prop:intercept}); and query-only resampling has coverage that tends to zero as queries are added when occasions carry shared effects (Proposition~\ref{prop:occasion}). A simulation study illustrates each result and the effect of extrapolating beyond the observed span (Section~\ref{sec:simulation}).
\item A ladder of five marginalizations between the cell quantity and a reported number, with the engine's query fan-out as the rung that no outside observer can condition on, and a decomposition showing how a change in fan-out appears as drift (Section~\ref{sec:ladder}).
\item An analysis of the admission stage (Section~\ref{sec:observability}): the nesting axiom behind the funnel factorization, the collider structure of the rendered list with a closed-form bias in a linear-Gaussian case, a description of what four public surfaces expose, the record of a study design we withdrew, and the identification strategies that remain.
\item A worked illustration on a preregistered panel of our own whose design could not identify the quantities it was built to measure (Section~\ref{sec:panel}), and a reporting checklist (Section~\ref{sec:checklist}).
\end{enumerate}

\paragraph{What this paper does not claim.} The estimands in this paper describe the observation process: draws, occasions, latent membership and their aggregation. None of them identifies the effect of an action taken by a publisher, by a client of ours, or by us, and we state the reason formally in Section~\ref{sec:limitations}. The authors run a company that sells measurement of visibility in AI answer engines; the conflict of interest is stated at the end of the paper.

\section{Setting: the observable and the measurement cell}
\label{sec:setting}

\paragraph{What is observed.} A measurement instrument for an answer engine records, for each request, the sources that the engine \emph{renders} on the answer surface: the reference list of an AI Overview, the citation list of a chat answer, or the grounding sources returned by an API. Nothing in such a record says which documents the engine retrieved, scored or read. We therefore define every estimand over a rendered/not-rendered indicator, and we avoid language about the engine's internal candidate set except where we define a latent quantity explicitly and state how it is or is not identified.

\paragraph{The cell.} Fix a query $q$ as issued by the user, a source $e$ at grain $L$ (URL, host, or registrable domain), and a capture path $\iota$, by which we mean the engine, its interface or API, the locale and device settings, the data provider if any, and the parser. Measurements are taken on occasions $t = 1, \dots, d$ at calendar times $\tau_1 < \dots < \tau_d$, and on each occasion by $k$ separate requests, which we call draws $j = 1,\dots,k$. The observable is
\begin{equation}
Y_{tj} = Y_{tj}(q,e,\iota) \in \{0,1\}, \qquad Y_{tj} = 1 \iff e \text{ is rendered as a source on draw } j \text{ of occasion } t,
\end{equation}
and $\cS_{tj} = \{e : Y_{tj} = 1\}$ is the rendered set at grain $L$. The quantity a repeated-query design can target directly is the cell rate
\begin{equation}
\theta(q,e,\iota,t) = \Prob\left(Y_{tj} = 1 \mid q, e, \iota, t\right).
\label{eq:cell}
\end{equation}

\paragraph{Grain.} A source at registrable-domain grain is rendered on a draw if any URL under that domain is rendered. Grain is a deterministic aggregation of the URL-level observable, and estimands at different grains are different estimands. They can disagree without either being wrong.

\paragraph{Missing outcomes.} A request can fail, or can return an answer without sources (for example, when the engine does not search). Whether such a draw contributes $Y_{tj}=0$ or is treated as missing changes the estimand: the first targets $\Prob(Y=1)$ and the second $\Prob(Y=1 \mid \text{answer with sources})$. Both are legitimate; they are not the same quantity, and a study must say which it reports. Scoring a capture failure as a non-render also adds spurious churn, since a failure on one occasion reads as an exit followed by a return.

\section{Measurement model and estimand family}
\label{sec:model}

\subsection{Latent membership and per-draw inclusion}

We separate two processes that the observable mixes. A latent membership indicator $M_t = M_t(q,e,\iota) \in \{0,1\}$ records whether, on occasion $t$, the source is among those the engine will render for the query at all. A per-draw inclusion probability $\pi = \pi(q,e,\iota)$ governs whether a member is rendered on a given draw. The model is
\begin{equation}
\Prob(Y_{tj} = 1 \mid M_t = 0) = 0, \qquad
\Prob(Y_{t1}=y_1, \dots, Y_{tk} = y_k \mid M_t = 1) = \prod_{j=1}^{k} \pi^{y_j}(1-\pi)^{1-y_j}.
\label{eq:model}
\end{equation}
The second statement is an assumption: draws on one occasion are conditionally independent given membership. It fails if, for example, a provider or engine caches answers, in which case same-occasion agreement overstates $\pi$. A design must defeat caching for this assumption to be plausible.

$M_t$ is defined by its consequence for rendering. It is not a claim that the engine maintains such a set. The same structure appears in ecological occupancy models, where a site is occupied or not and a species present at an occupied site is detected with probability below one \citep{mackenzie2002occupancy}, and its dynamic version with colonization and local extinction \citep{mackenzie2003dynamic,royle2007dynamic}. The identification results of Section~\ref{sec:identification} are, in that vocabulary, the statement that repeat visits within a period of closure are needed to separate detection from occupancy. We adopt the analogy because the identification arguments carry over directly.

\paragraph{Observed rotation overstates latent rotation.} Under~\eqref{eq:model}, a member that is not drawn on one occasion and drawn on the next looks like an exit and an entry. A hidden Markov deconvolution \citep{rabiner1989hmm} of a panel of Google AI Overview captures (4{,}471 series, one per query and domain, over 8 waves in July 2026) turned a raw per-wave admission exit rate of $0.457$ ($3{,}506/7{,}673$ transitions) into a latent exit rate of $0.0445$, 95\% CI $[0.033, 0.059]$, using an emission model fitted on a separate panel with same-day redraws. The deconvolved figure depends on that emission model, and our internal review of it is not complete; the qualitative point, that most observed rotation is draw noise, is what the model predicts whenever $\pi$ is well below one.

\subsection{Estimands for citation stability}

All estimands below are defined for a fixed grain $L$ and capture path $\iota$ and then averaged over a declared population of queries. We write them at the set level, where most studies report, and note that each is a weighted average of entity-level conditional probabilities.

\begin{definition}[Same-occasion noise floor]
$N_L = \E\left[\,|\cS_{t1} \cap \cS_{t2}| \,/\, |\cS_{t1}|\,\right]$, the expected share of the sources rendered on one draw that are rendered again on a second draw of the same occasion. At entity level and under~\eqref{eq:model}, $\Prob(Y_{t2}=1 \mid Y_{t1}=1) = \pi$.
\end{definition}

\begin{definition}[Across-occasion retention]
$R_L(\Delta; u) = \E\left[\,|\cS_t^{u} \cap \cS_{t'}^{u}| \,/\, |\cS_t^{u}|\,\right]$ for occasions with $\tau_{t'} - \tau_t = \Delta$, where the union convention $u$ states whether $\cS_t^{u}$ is a single draw or the union of the $k$ draws of occasion $t$. Retention is asymmetric (it depends on which occasion is the anchor). The Jaccard index $|\cS \cap \cS'|/|\cS \cup \cS'|$ is a different, symmetric and smaller quantity.
\end{definition}

The union convention matters because, under~\eqref{eq:model}, a member enters the union of $k$ draws with probability $1-(1-\pi)^k$, which increases in $k$. Unioning inflates every set and every overlap computed from it, so overlap statistics with different $k$ are not comparable even when they use the same metric.

\begin{definition}[Stable core as a threshold family]
For a panel of $d$ occasions, $K_c(q) = \{e : \sum_{t} \indic[e \in \cS_t^{u}] \geq c\}$ for $c = 1,\dots,d$, reported as the whole curve $c \mapsto (|K_c|, \text{coverage of } K_c)$, where coverage is the share of a typical occasion's rendered set that lies in $K_c$. A single threshold reported alone is a choice presented as a measurement.
\end{definition}

\begin{definition}[Recurrence response]
\[
\rho_m(h; H) = \Prob\left(e \in \cS_H^{u} \mid e \text{ rendered on exactly } h \text{ of the first } m \text{ occasions}\right),
\]
for $h = 1,\dots,m$ and a later occasion $H$. It is model-free and in-sample when $H$ lies within the observed span.
\end{definition}

\subsection{The funnel and its nesting axiom}

Studies of the citation funnel write the probability of being cited as a product of stages \citep{martinez2026survey}. Let $G \in \{0,1\}$ indicate that a source is \emph{admitted}, meaning that it is among the sources from which the answer is generated, and let $C \in \{0,1\}$ indicate that it is cited. The terms ``retrieval pool'' and ``admitted set'' name the same set; we use \emph{admitted} throughout. The factorization
\begin{equation}
\Prob(C=1) = \Prob(C=1 \mid G=1)\,\Prob(G=1)
\label{eq:funnel}
\end{equation}
holds if and only if $\Prob(C=1, G=0) = 0$, that is, if every cited source was admitted. We call this the \emph{nesting axiom}. It is plausible for the latent object $G$. It is a testable property of any observed proxy $\tilde G$, and it can fail.

\begin{proposition}[Products of proxy stages]
\label{prop:nesting}
For any binary proxy $\tilde G$,
$\Prob(C=1 \mid \tilde G=1)\,\Prob(\tilde G=1) = \Prob(C=1) - \Prob(C=1, \tilde G=0)$.
\end{proposition}

\begin{proof}
The left side equals $\Prob(C=1,\tilde G=1)$, and $\Prob(C=1) = \Prob(C=1,\tilde G=1) + \Prob(C=1,\tilde G=0)$.
\end{proof}

A product of stage rates computed from a proxy that violates nesting understates the citation rate by the citation mass outside the proxy, and the stage-wise ``conditional citation rate'' is then a ratio over two sets that are not nested, not a conditional probability of citation given admission. Section~\ref{sec:observability} shows that this is the situation on one of the surfaces we have captured.

\subsection{The universe and the collapse rule are part of the estimand}

Two definitional choices that are often left implicit move reported numbers by large factors.

\emph{The universe.} A share of explained variation in $\Prob(C=1)$ depends on which (query, source) cells are counted. If the universe includes sources that were never admitted and therefore have $C=0$ by construction, part of any ``admission share'' is explained by the construction itself. In a deviance decomposition on Google AI Overview captures (13{,}581 cells defined by domain, query and wave; 8 waves; 150 queries; July 2026), the share attributed to the admission stage was $0.773$, 95\% CI $[0.746, 0.800]$, with a universe in which $34.4\%$ of cells were never admitted, and $0.613$, 95\% CI $[0.540, 0.673]$, with a stricter definition that avoids the deterministic zeros. Neither number describes leverage or effect; both describe how outcome variation is partitioned under a stated universe, and they should be reported together.

\emph{The collapse rule.} A state defined over several occasions (``admitted'', ``cited'') requires a rule that maps $W$ occasion-level observations to one label. The rule \emph{rendered on any of $W$ occasions} targets $\Prob(\exists t \le W: Y_t = 1)$, which is non-decreasing in $W$ and, when render events on different occasions are independent with probabilities $\theta_t$, equals $1 - \prod_{t \le W}(1-\theta_t)$. It is a property of the window, not of any occasion. On the same Google AI Overview panel (5{,}950 cells defined by query and domain, over 142 queries), the any-occasion rule over 11 to 13 waves labelled $4{,}266/5{,}950 = 0.717$ of cells as admitted under the render proxy, against a mean single-wave rate of $0.193$ and a rate of $1{,}959/5{,}950 = 0.329$ under a rule requiring at least two occasions (Wilson intervals, which ignore clustering by query, $[0.705, 0.728]$ and $[0.317, 0.341]$). Section~\ref{sec:simulation} reproduces the shape of this dependence on synthetic data. A definition moved the headline by a factor of more than three; the estimand was not determined until the rule was named.

\section{Identification by design}
\label{sec:identification}

This section states what repeated-query designs can and cannot identify. The results are elementary, and that is part of the point: each of them is violated by common designs.

\subsection{Draw noise and fast churn require \texorpdfstring{$k \geq 2$}{k >= 2}}

Generalize~\eqref{eq:model} by adding an occasion-level factor. Let $(S_t)$ be a stationary two-state Markov chain (slow membership), let $F_t \sim \mathrm{Bernoulli}(c)$ independently across occasions and independent of $(S_t)$ (a fast component that refreshes on every occasion), let $M_t = S_t F_t$, and let draws be as in~\eqref{eq:model} with inclusion probability $\pi$.

\begin{proposition}[Identification of the noise floor]
\label{prop:k}
(a) With $k=1$ draw per occasion, the joint distribution of $(Y_{11}, \dots, Y_{d1})$ depends on $(\pi, c)$ only through the product $\pi c$, for every $d$. Hence $\pi$ and $c$ are not separately identified, by any number of occasions.\\
(b) With $k \geq 2$ draws per occasion and the conditional independence in~\eqref{eq:model}, $\Prob(Y_{t2}=1 \mid Y_{t1}=1) = \pi$, so $\pi$ is identified. Given $\pi$, the single-draw retention curve $\Prob(Y_{t+\Delta,1}=1 \mid Y_{t1}=1) = \pi c\, \Prob(S_{t+\Delta}=1 \mid S_t=1)$, $\Delta \geq 1$, then identifies $c$ and the parameters of a non-degenerate chain.
\end{proposition}

\begin{proof}
(a) With one draw, $Y_{t1} = S_t \cdot (F_t B_{t1})$, where $B_{t1} \sim \mathrm{Bernoulli}(\pi)$. The variables $H_t = F_t B_{t1}$ are independent across $t$, independent of $(S_t)$, and $\mathrm{Bernoulli}(\pi c)$. The law of $(Y_{t1})_t$ is therefore a function of the law of $(S_t)$ and of $\pi c$ only.
(b) $\Prob(Y_{t1}=1, Y_{t2}=1) = \Prob(S_t=1)\, c\, \pi^2$ and $\Prob(Y_{t1}=1) = \Prob(S_t=1)\, c\, \pi$; the ratio is $\pi$. For a stationary two-state chain, $\Prob(S_{t+\Delta}=1 \mid S_t=1) = \mu + (1-\mu)\lambda^{\Delta}$ with occupancy $\mu$ and autocorrelation $\lambda$, so the retention curve has the form $a + b\lambda^{\Delta}$ with $a = \pi c \mu$ and $b = \pi c (1-\mu)$; with $\pi$ known, $c = (a+b)/\pi$ and $\mu = a/(a+b)$.
\end{proof}

The proposition says that a single-draw design cannot tell an engine that re-samples sources on each request from one whose membership changes every day. Both produce the same retention curve (Figure~\ref{fig:sim}a). A model with no fast component ($c=1$) identifies $\pi$ by extrapolating the retention curve to lag zero, but that identification rests entirely on the parametric form. The assumption excludes exactly the alternative a skeptic would raise. In the vocabulary of occupancy models, $k \ge 2$ is the requirement of repeat visits within a closed period \citep{mackenzie2002occupancy}.

The practical consequence is that the noise floor must be measured, not assumed, and that the number of draws per occasion is the first design parameter to fix. In a panel of ChatGPT captures through the DataForSEO scraper of the ChatGPT consumer interface (100 queries, 12 occasions over 6 days, 3 draws per occasion, July 2026), the Jaccard agreement of cited-domain sets was $0.304$, 95\% CI $[0.293, 0.315]$, between draws of one occasion ($n = 3{,}600$), $0.297$ $[0.288, 0.306]$ across occasions on the same day, and $0.279$ $[0.276, 0.282]$ across days; a nested variance decomposition of binary domain presence attributed $66\%$ of the variance to within-occasion redraws and under $4\%$ to day and occasion effects combined (no interval is recorded for the variance shares). An independent study of four engines reached the same qualitative conclusion: same-day re-runs agreed about as little as consecutive days \citep{schulte2026dontmeasureonce}. Both are designs with $k \geq 2$, and only such designs can make this comparison.

\paragraph{Small $k$.} With $k$ draws and no render, the exact one-sided 95\% upper bound for a per-draw probability is $1 - 0.05^{1/k}$, which is $0.63$ for $k=3$ and close to $3/k$ for larger $k$. Zero renders in three draws is compatible with a per-draw rate above one half, and a single non-render is compatible with any rate up to $0.95$. A single-score check of a source for a query cannot distinguish a source that is usually rendered from one that rarely is.

\subsection{Normalizing by the noise floor does not make a decay fit conservative}

A common analysis divides across-occasion retention by the noise floor, fits a log-linear decay in the lag, and projects to a longer horizon. It is sometimes argued that dividing by an imperfectly measured floor biases the projection toward the null.

\begin{proposition}[Free-intercept invariance]
\label{prop:intercept}
Let $\hat\beta$ be the ordinary least squares slope of $\log \hat R(\Delta)$ on $\Delta$ with a free intercept, over any set of lags, and let $\hat\beta_N$ be the slope of $\log\left(\hat R(\Delta)/N\right)$ on $\Delta$ for any constant $N>0$. Then $\hat\beta_N = \hat\beta$, and every ratio of projections $\hat R(H)/\hat R(H')$, including any half-life, is the same under both fits.
\end{proposition}

\begin{proof}
$\log(\hat R/N) = \log \hat R - \log N$; subtracting a constant from the response changes only the intercept of a least squares fit with a free intercept.
\end{proof}

The normalization therefore changes the level of a projection and nothing about its rate. Any claim that it makes the rate conservative is false for this model class. We report this because a preregistration of ours made the claim (Section~\ref{sec:panel}).

\subsection{Shared occasions: what resampling queries cannot see}

Retention between occasions $s$ and $t$ is computed from the same two occasions for every query in a panel, so anything specific to those occasions (an engine update, a provider incident, a news event) is shared by all queries.

\begin{proposition}[Occasion variance]
\label{prop:occasion}
Let the retention of query $q$ between occasions $s<t$ be $R_{q,s,t} = f(\tau_t - \tau_s) + u_s + u_t + \varepsilon_{q,s,t}$, with occasion effects $u_t$ independent with mean zero and variance $\sigma_u^2$, and errors $\varepsilon$ independent of $u$ with mean zero and variance $\sigma_\varepsilon^2$. Let $P_\ell$ be the set of occasion pairs at lag $\ell$ and $\hat f(\ell)$ the mean of $R_{q,s,t}$ over $n$ queries and the pairs in $P_\ell$. Then
\begin{equation}
\Var\left(\hat f(\ell)\right) = \frac{\sigma_u^2}{|P_\ell|^2}\sum_{t} m_t^2 + \frac{\sigma_\varepsilon^2}{n|P_\ell|} \;\geq\; \frac{2\sigma_u^2}{|P_\ell|} + \frac{\sigma_\varepsilon^2}{n|P_\ell|},
\label{eq:occvar}
\end{equation}
where $m_t$ is the number of pairs in $P_\ell$ that contain occasion $t$. The first term does not shrink with $n$. A bootstrap that resamples queries with occasions held fixed estimates only the second term, so, for Gaussian $u$ and $\sigma_u>0$, the coverage of its nominal intervals tends to zero as $n \to \infty$.
\end{proposition}

\begin{proof}
$\hat f(\ell) - f(\ell) = |P_\ell|^{-1}\sum_{(s,t)\in P_\ell}(u_s+u_t) + \bar\varepsilon$, and the occasion sum equals $\sum_t m_t u_t$. Since $\sum_t m_t = 2|P_\ell|$ and $m_t^2 \geq m_t$ for non-negative integers, $\sum_t m_t^2 \geq 2|P_\ell|$. Conditional on $u$, the only variation across query resamples is in $\bar\varepsilon$, whose variance is $\sigma_\varepsilon^2/(n|P_\ell|)$. The interval half-width therefore tends to zero while the estimation error converges in distribution to a non-degenerate normal variable, so the probability that the error lies inside the interval tends to zero.
\end{proof}

Two design rules follow. Intervals for across-occasion quantities must resample occasions as well as queries, for example with a two-way bootstrap for crossed designs \citep{owen2007pigeonhole} or multiway clustering \citep{cameron2011multiway}, and a lag represented by one occasion pair has no estimable between-occasion variance at all. Second, short panels concentrate their longest lags on few pairs that share anchor occasions: with occasions at days $0, 2, 3, 6, 7, 8$, the lags of 7 and 8 days are each represented by a single pair, both anchored on the first day, so any peculiarity of that day is indistinguishable from an effect of lag. The span of a panel must be a multiple of the horizon of interest, with several pairs at each lag that do not share an anchor.

Serial correlation between repeated observations of the same query reduces the information in a long series further. Under first-order autocorrelation $\rho$, $n$ observations carry roughly the information of $n(1-\rho)/(1+\rho)$ independent ones \citep{bayley1946effective}. We print the formula without a value of $\rho$: it should be estimated within the panel being analysed.

\subsection{Extrapolation beyond the observed span}

If some sources persist indefinitely, retention approaches a positive plateau, and a model such as $R(\Delta) = a + b\,e^{-\kappa\Delta}$ with $a>0$ is appropriate. A log-linear model with no asymptote must then absorb the plateau into its rate. For any finite span, its projection $\exp(\hat\alpha + \hat\beta H)$ tends to zero as $H$ grows while the true value tends to $a$, so for long enough horizons the no-asymptote projection is biased downward. The asymptote model avoids this bias but its parameter $a$ is weakly identified on short spans, where slow decay and fast decay to a plateau fit the same points. Section~\ref{sec:simulation} quantifies both effects. A reported horizon beyond the observed span should carry an explicit extrapolation flag.

\section{A simulation study}
\label{sec:simulation}

We illustrate Sections~\ref{sec:model} and~\ref{sec:identification} on synthetic panels generated from the model of Proposition~\ref{prop:k}. Latent membership is a stationary two-state Markov chain with occupancy $0.35$ and lag-one autocorrelation $0.9$, so latent retention is $\Prob(S_{t+\Delta}=1 \mid S_t=1) = 0.35 + 0.65 \cdot 0.9^{\Delta}$. Every number in this section is synthetic; none comes from an engine. The code is the ancillary file \texttt{anc/simulation.py} (fixed seed), and all values below are written by it to \texttt{simulation\_summary.json}.

\begin{figure}[t]
\centering
\includegraphics[width=\textwidth]{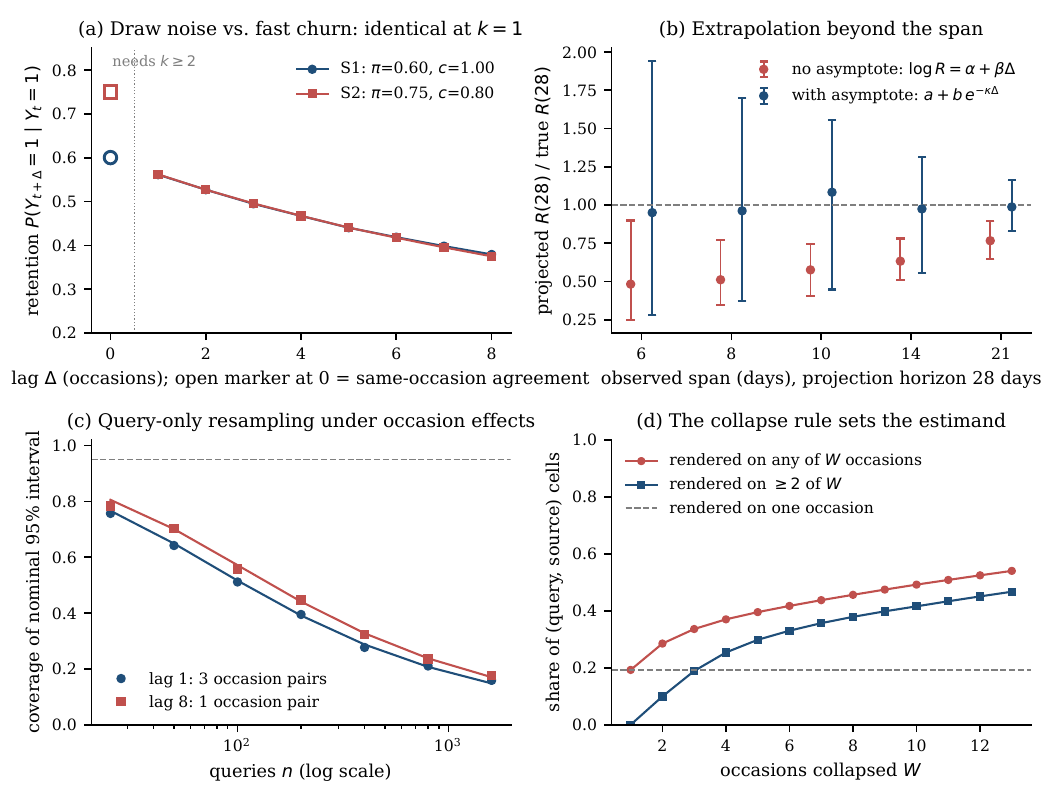}
\caption{Simulation study on synthetic data. (a) Two parameter settings with equal $\pi c$ give the same single-draw retention curve; only the same-occasion agreement (open markers, which require $k \geq 2$) separates them (Proposition~\ref{prop:k}). (b) Ratio of the projected to the true 28-day retention, median and 2.5 to 97.5 percentile over 200 replicate panels of 1{,}500 entities with one draw per daily occasion. (c) Coverage of nominal 95\% query-only intervals under occasion effects ($\sigma_u = 0.04$, $\sigma_\varepsilon = 0.25$), with occasions at days $0, 2, 3, 6, 7, 8$; points are 2{,}000 simulated panels per value of $n$, lines are the closed form implied by Proposition~\ref{prop:occasion}. (d) Share of cells labelled ``rendered'' under three collapse rules as the number of collapsed occasions grows ($\pi = 0.55$).}
\label{fig:sim}
\end{figure}

\paragraph{(a) Equivalence at $k=1$.} Setting S1 has $\pi = 0.60$ and no fast component ($c = 1$); setting S2 has $\pi = 0.75$ and $c = 0.80$. With 200{,}000 entities, their single-draw retention at lag one is $0.561$ and $0.562$, and at lag eight $0.379$ and $0.375$; the theoretical value is $0.378$ for both. Their same-occasion agreement, which requires a second draw, is $0.600$ and $0.750$.

\paragraph{(b) Extrapolation.} We fit both models of Section~\ref{sec:identification} to single-draw retention over spans of 6 to 21 days and project to 28 days (true value $0.230$). The log-linear projection is biased downward at every span: its median ratio to the truth is $0.48$ at a 6-day span and $0.77$ at 21 days, and its 97.5th percentile is below one at every span. The asymptote model is centred near the truth (median ratio $0.95$ to $1.08$) but is unusable on short spans: its central 95\% range is $[0.28, 1.94]$ at 6 days and narrows to $[0.83, 1.16]$ at 21 days. A short panel can choose between a biased projection and an uninformative one.

\paragraph{(c) Coverage.} With the occasion layout above, nominal 95\% query-only intervals cover the true lag effect in $76\%$ of panels with 25 queries and $16\%$ with 1{,}600 queries at lag one (three occasion pairs), and $78\%$ and $18\%$ at lag eight (one pair). Adding queries makes these intervals more confident and less often correct.

\paragraph{(d) Collapse rules.} The per-occasion render rate is $0.19$. After 12 occasions, the any-occasion rule labels $0.53$ of cells as rendered and the at-least-two rule $0.45$; both rise with $W$ while nothing about any single occasion has changed. The empirical pattern in Section~\ref{sec:model} has the same shape.

The study substitutes evidence of recovery for a proof of identifiability under general budgets. We have not characterized, for general $(k, d, n)$ and general latent dynamics, when window length and per-draw inclusion are separately identified; we state the question as open in Section~\ref{sec:limitations}.

\section{The marginalization ladder and query fan-out}
\label{sec:ladder}

The cell rate~\eqref{eq:cell} is the most conditional quantity a repeated-query design can target. Reported numbers are obtained from it by up to five marginalizations, which we list from the inside out.

\begin{enumerate}
\item[\textbf{R1.}] \textbf{Draws.} A per-draw rate $\theta$ versus the rate of appearing in the union of $k$ draws, $1-(1-\pi)^k$ for a member. The union convention and $k$ must be declared.
\item[\textbf{R2.}] \textbf{Occasions.} Averaging $\theta$ over a window of occasions, or collapsing the window with a rule such as ``any'' or ``at least $c$'' (Section~\ref{sec:model}). The window and the rule must be declared.
\item[\textbf{R3.}] \textbf{Queries.} A statement about a source, $\bar\theta_w(e) = \sum_q w(q)\,\theta(q,e,\iota,t)$, averages over a population of queries with weights $w$. The population is usually a convenience panel, is rarely stated, and is not the population that matters to a given reader.
\item[\textbf{R4.}] \textbf{Capture paths.} Pooling across engines, interfaces, locales or providers. Different surfaces return substantially different sources for the same query \citep{grossman2026disrupts}, so a pooled rate describes the mixture of paths in the sample and no engine in particular.
\item[\textbf{R5.}] \textbf{The engine's query fan-out.} Conditioning on the user query $q$ leaves a marginalization that the engine performs.
\end{enumerate}

Rung R5 needs a formal statement. Suppose the engine issues a random set of sub-queries $\cQ \sim F_{\iota t}(\cdot \mid q)$ before or while it generates an answer, and let $h_{\iota t}(\cQ, e)$ be the probability that $e$ is rendered given the realized sub-query set. By the law of total probability,
\begin{equation}
\theta(q,e,\iota,t) = \E_{\cQ \sim F_{\iota t}(\cdot \mid q)}\left[h_{\iota t}(\cQ, e)\right].
\label{eq:fanout}
\end{equation}
The cell rate is therefore a mixture over a distribution that the engine owns, that is not observed in full, and that can change from one occasion to the next. Writing $\E_t$ for the expectation under $F_{\iota t}$, the change in the cell rate between occasions decomposes as
\begin{equation}
\theta_{t+1} - \theta_t = \underbrace{\left(\E_{t+1} - \E_{t}\right)\left[h_{t+1}\right]}_{\text{change in fan-out}} + \underbrace{\E_{t}\left[h_{t+1} - h_{t}\right]}_{\text{change given fan-out}}.
\label{eq:drift}
\end{equation}
A change in how the engine expands a query appears as drift in the cell, on exactly the occasion axis that stability studies measure, and no outside design can separate the two terms without observing $\cQ$.

Some surfaces expose part of $\cQ$. Among the ChatGPT captures we hold that expose a sub-query list (90 panels, 30 queries with 3 draws each, July 2026), the list had one entry on 85 panels, two on 4 and three on 1; among 179 Gemini API responses with search grounding (July 2026), 108 exposed no search queries, 10 exposed one, 54 exposed two and 7 exposed three. These lists should be read as lower bounds on expansion: on the ChatGPT surface most credited sources lie outside the search results the surface exposes (Section~\ref{sec:observability}), so the exposed sub-queries and their results do not account for most of the sources the answer credits, which came from retrieval the surface does not show or from no retrieval at all. Engines appear to disclose the breadth of their search more readily than the evidence they used, and neither disclosure is complete.

\paragraph{A worked example from our own reporting.} A statement cleared for our own public reporting reads: ``Across the non-branded searches we measured, more than half of the sites AI Overviews cited weren't in Google's top 10. Good rankings don't guarantee your page is the one being read.'' The underlying quantity is a pooled share of cited registrable domains that are absent from the same page's organic top ten, on Google AI Overviews for a panel of non-branded US English searches. In the terms of this section it is a marginal over draws and occasions (pooled waves, R1 and R2), over queries (a fixed panel of five commercial verticals, R3), and over the engine's fan-out (R5), at one capture path (R4 held fixed) and one grain. It is a statement about that panel. It is not a statement about any one query or site, and it does not say that the cited sites outside the top ten were retrieved by some other route; it says only what was rendered.

\section{Observability of the admission stage}
\label{sec:observability}

The funnel factorization~\eqref{eq:funnel} is useful only if admission $G$ can be observed or identified. This section asks what an admission instrument would have to observe, argues that the rendered list is a collider rather than a noisy proxy, reports which fields four public surfaces expose, records a design we withdrew, and describes the identification strategies that remain.

\subsection{What an admission instrument must observe}

An instrument for $G$ must produce an observation that depends on admission and is not itself an effect of the answer. A list that the engine writes together with the answer (a reference block, a citation list, a set of grounding sources attached to answer spans) satisfies the first requirement and fails the second. Such a list is attribution: it records which sources the engine chose to credit, after or while it chose what to say.

\subsection{The rendered list as a collider}

\begin{figure}[t]
\centering
\begin{tikzpicture}[node distance=12mm and 18mm, every node/.style={draw, rounded corners, minimum width=11mm, minimum height=7mm, font=\small}, >={Stealth[length=2.2mm]}]
\node (G) {$G$};
\node (Gt) [right=of G] {$\tilde G$};
\node (U) [right=of Gt] {$U$};
\node (Y) [below=of Gt] {$Y$};
\draw[->] (G) -- (Gt);
\draw[->] (U) -- (Gt);
\draw[->] (U) -- (Y);
\draw[->] (G) -- (Y);
\end{tikzpicture}
\caption{Admission $G$, the rendered source list $\tilde G$, answer-level factors $U$, and an answer-level outcome $Y$. Conditioning on $\tilde G$, the collider, opens a non-causal path between $G$ and $Y$ through $U$.}
\label{fig:dag}
\end{figure}
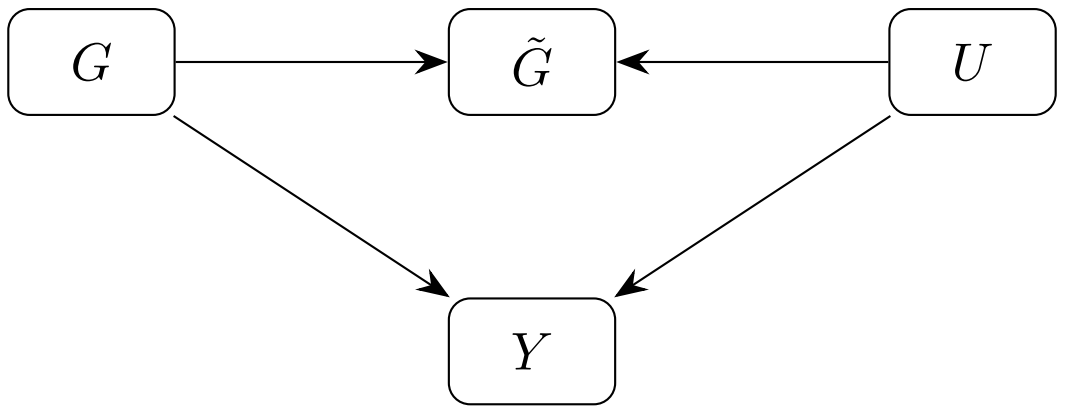

Let $\tilde G$ indicate that a source appears anywhere in the rendered source list, the usual proxy for admission, and let $Y$ be an answer-level outcome of interest, for example that the source is cited at the top level or quoted. Let $U$ collect query- and answer-level factors (the form of the answer, its length, the claims it makes) that influence both what the answer says and how its sources are displayed. The engine decides the layout jointly with the content, which gives the structure in Figure~\ref{fig:dag}: $\tilde G$ is a common effect of $G$ and $U$, and $U$ also affects $Y$.

Conditioning on a common effect induces an association between its causes \citep{berkson1946,greenland2003collider,cole2010collider,pearl2009causality}. Every estimate of the form ``$Y$ among sources that appear in the rendered list'' conditions on $\tilde G$. A linear-Gaussian version makes the size of the problem explicit.

\begin{proposition}[Collider bias in a linear-Gaussian model]
\label{prop:collider}
Let $G$ and $U$ be independent standard normal variables, $\tilde G = G + U + \eta$ with $\eta \sim N(0,\sigma^2)$ independent, and $Y = \beta G + \gamma U + \epsilon$ with $\epsilon$ independent noise. The coefficient on $G$ in the regression of $Y$ on $(G, \tilde G)$ is
\begin{equation}
\beta - \frac{\gamma}{1+\sigma^2}.
\end{equation}
\end{proposition}

\begin{proof}
For jointly Gaussian variables the regression is $\E[Y \mid G, \tilde G] = \beta G + \gamma\,\E[U \mid G, \tilde G]$. Given $G$, $\tilde G - G = U + \eta$, so $\E[U \mid G, \tilde G] = (\tilde G - G)/(1+\sigma^2)$, whose coefficient on $G$ is $-1/(1+\sigma^2)$.
\end{proof}

Three properties of this bias distinguish it from measurement error. It does not shrink with sample size, because it is a property of the conditioning set. It does not shrink with better extraction: a perfect parser measures $\tilde G$ exactly, and $\sigma^2 = 0$ is the case of largest bias. And it can reverse a sign, whenever $\gamma/(1+\sigma^2) > \beta$. The proposition is an illustration of the mechanism, not a model of any engine; binary versions with selection on $\tilde G = 1$ behave the same way qualitatively \citep{elwert2014endogenous}.

The structure also bears on latent-state models that treat the rendered list as an emission from a hidden admission state, $\Prob(\tilde G \mid G)$. That factorization presumes that the readout is generated from the state alone. If $U \to \tilde G$, the factorization is the assumption under dispute, and fitting it more carefully does not test it. When the exposed readout is post-selection by construction, the emission model is degenerate and no amount of fitting recovers the state.

What would refute the collider reading is evidence that rendering decisions depend on admission alone, that is, $\tilde G \perp U \mid G$. This is an empirical question, and it is the question an admission instrument would answer.

\subsection{What four public surfaces expose}

We examined captured responses from four surfaces for any field that records a set of sources larger than, and containing, the set the answer credits. Table~\ref{tab:surfaces} summarizes them.

\begin{table}[t]
\centering
\small
\caption{Source fields exposed by four public AI answer surfaces (July 2026 captures).}
\label{tab:surfaces}
\setlength{\tabcolsep}{5pt}
\renewcommand{\arraystretch}{1.15}
\begin{tabular}{@{}>{\raggedright\arraybackslash}p{4.1cm}>{\raggedright\arraybackslash}p{2.85cm}>{\raggedright\arraybackslash}p{3.25cm}rrr@{}}
\toprule
Surface & Candidate field & Credited field & Responses & Candidates & \begin{tabular}[b]{@{}r@{}}Not\\credited\end{tabular} \\
\midrule
Perplexity \newline {\footnotesize Sonar API} & \texttt{search\_results} & \texttt{citations} & 77 & 1{,}150 & 0 \\[2pt]
Gemini \newline {\footnotesize API with search grounding} & \texttt{groundingChunks} & chunks cited in \texttt{groundingSupports} & 179 & 741 & 0 \\[2pt]
Google AI Overviews \newline {\footnotesize DataForSEO SERP API} & none (one list only) & reference list & 1{,}710 & 10{,}125 & n/a \\[2pt]
ChatGPT \newline {\footnotesize DataForSEO scraper of the consumer interface, web search forced} & \texttt{search\_results} & \texttt{sources} & 180 & 1{,}481 & 1{,}476 \\
\bottomrule
\end{tabular}

\smallskip
\begin{minipage}{\linewidth}\footnotesize
\emph{Candidate field}: the larger field that might be read as a retrieval pool. \emph{Credited field}: the field the answer cites. \emph{Not credited}: candidates absent from the credited field. Model identities as reported in the responses: Perplexity \texttt{sonar}, Gemini \texttt{gemini-3.6-flash}, ChatGPT \texttt{gpt-5-5} (as reported by the capture provider). Client-audit captures, and the captures of one internal study whose results are not yet released, are excluded.
\end{minipage}
\end{table}

On the Perplexity and Gemini surfaces, every candidate in every captured response is credited: the two fields coincide, so the middle state of the funnel (admitted but not credited) is empty in all responses we hold. The AI Overview surface exposes a single list. On the ChatGPT surface the two fields differ, but they are not nested: of 604 credited URLs, 5 appear among the 1{,}481 URLs in the exposed search results. That field is a second rendered list, plausibly the visible result set of one search, and not a pool from which the credited sources were selected. Proposition~\ref{prop:nesting} applies directly: a ``conditional citation rate'' computed from these two fields is a ratio of the sizes of two non-nested sets.

The pattern is what one would expect if the exposed provenance is attribution produced with the answer. Google's own description of AI features is consistent with this: ``While responses are being generated, our advanced models identify more supporting web pages'' \citep{googleaifeatures}. We do not know whether the coincidence of fields on the Perplexity surface is a property of the API or of the queries we sent; a probe designed to force divergence (queries with broad result sets and narrow answers) would settle it, and we have not run one.

\subsection{A withdrawn design}

An earlier version of this work proposed to measure the collider directly with two engines: Perplexity's \texttt{search\_results} would observe admission, Google's reference list would observe only the render, and the gap on matched queries would measure how much of the rendered middle state was presentation rather than grounding. The premise was that \texttt{search\_results} is a retrieval field. In every captured Perplexity response we hold, it is identical to the citation list, so the design has no admission arm and we have withdrawn it.

We record the withdrawal because of how the error arose. The study that first captured the Perplexity responses recorded, among its limitations, that the two fields coincided on every row. Later internal documents cited that study for the existence of a retrieval field and left the limitation behind. A field's name is not a measurement of what it contains; a claim that a field is a retrieval pool, a render or a citation list needs a computation over rows.

\subsection{A pilot test of render dependence}

We ran one pilot on data we already held. On 26 queries captured on both ChatGPT (web search forced) and Google AI Overviews, we took the hosts in ChatGPT's exposed search results and compared the AI Overview render status of hosts that ChatGPT credited with that of hosts at the nearest rank that it did not credit. Across the 18 queries in which both groups occurred, the mean difference in AI Overview render probability was $0.28$, 95\% bootstrap interval over queries $[0.11, 0.50]$; a negative control that assigned the grouping by a hash of the URL gave $-0.04$ $[-0.14, 0.06]$.

This is a pilot, and it cannot produce a verdict. The sample is 18 query cells. The AI Overview surface exposed no source that was listed and not credited, so on the Google side the render and the credit coincide and the pilot cannot separate the two nodes. The association is consistent with answer-level factors shared across engines that influence rendering, but it is equally consistent with the two engines retrieving similar sources, and the comparison crosses engines. What the pilot establishes is that the instrument runs and produces a non-degenerate contrast against a negative control.

\subsection{Identification strategies that remain}

On the surfaces we have measured, admission is not identified from engine-side observation. Three routes remain.

\emph{Controlled apparatus.} When the experimenter sets the context passed to a model, admission is known by construction and the second stage, $\Prob(C \mid G)$, is identified for that apparatus. This is the design of much of the optimization literature \citep{aggarwal2024geo,puerto2025cseo,vishwakarma2026whatgetscited}. It identifies an effect conditional on a pinned context and does not by itself transfer to a live engine, whose retrieval, context allocation and layout are not pinned \citep{martinez2026survey}.

\emph{Publisher-side observation.} Where an engine fetches pages at answer time, a publisher's server log records a fetch that precedes the render and that the answer did not write. With the consent of site owners, such logs can observe an event close to admission for their own pages. The limits are real: not every engine fetches at answer time, a fetch is not use, and coverage is limited to consenting sites.

\emph{Contract tests on API surfaces.} Probes designed to make exposed fields diverge can establish whether a field is a pool or an attribution list on a given surface. They cost little and should precede any study that treats a field as a pool.

Until one of these is in place, stage-wise estimates computed from rendered lists should be reported as render-conditional quantities, with the conditioning stated.

\section{Illustration: a preregistered design that could not identify its target}
\label{sec:panel}

This section applies Sections~\ref{sec:model} and~\ref{sec:identification} to a panel of our own. We report it as a worked example of a design that failed identification, not as a source of stability values.

\paragraph{Design.} A panel of 100 non-branded queries in five commercial verticals was captured on Google AI Overviews (US English, desktop, top ten organic results) through the DataForSEO SERP API, on six collection days spanning eight calendar days in July 2026 (days $0, 2, 3, 6, 7, 8$), with three same-day draws per day: 1{,}800 requests. Of the 100 queries, 96 returned an AI Overview reference list on at least one day and 91 on all six days. The estimands were preregistered before analysis: the noise floor, retention by lag at URL and domain grain, a log-linear projection to 28 days, and a stable core at a fixed threshold. A preregistered decision rule compared the projection with a floor taken from the specification of our own product, which at the time named individual URLs as objects that would persist over a client's action horizon.

\begin{figure}[t]
\centering
\includegraphics[width=0.85\textwidth]{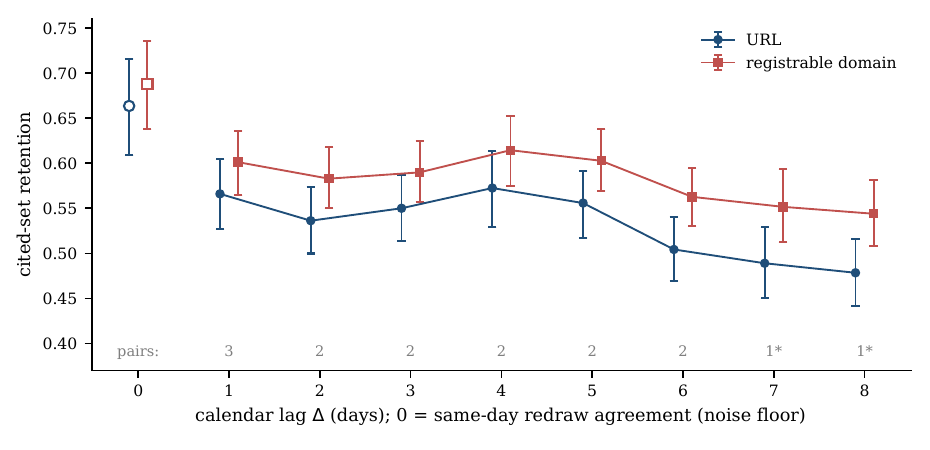}
\caption{Cited-set retention by calendar lag on the preregistered Google AI Overview panel (Section~\ref{sec:panel}), with query-cluster bootstrap 95\% intervals. Open markers at lag 0 are the same-day noise floor. The numbers above the axis are the occasion pairs available at each lag; an asterisk marks lags whose only pair is anchored on the first collection day. The intervals resample queries only and omit between-occasion variance (Proposition~\ref{prop:occasion}).}
\label{fig:ladder}
\end{figure}

\paragraph{Verdict.} The rule returned its failing outcome at both grains (the preregistered label was ``stale on arrival''). The noise floor was $0.664$, 95\% CI $[0.609, 0.716]$, at URL grain and $0.688$ $[0.638, 0.736]$ at domain grain (query-cluster bootstrap, 2{,}000 draws): two identical requests on the same day agreed on about two thirds of the cited URLs. Retention at the longest observed lag was $0.478$ $[0.441, 0.516]$ at URL grain and $0.544$ $[0.508, 0.581]$ at domain grain (Figure~\ref{fig:ladder}). The product specification was changed as a result.

\paragraph{What the design could not identify.} An independent re-analysis of the panel found four defects, each an instance of Section~\ref{sec:identification}.
\begin{enumerate}
\item \emph{Day identity and lag were confounded.} The first collection day had lower retention against every other day at every lag. In a pair-level regression over the 15 day pairs, adding an indicator for pairs anchored on the first day reduced the URL lag coefficient by 66\% (from $-0.0104$ to $-0.0035$ per day) and the domain coefficient by 97\% (from $-0.0064$ to $-0.0002$ per day). Dropping the first day reduced the $R^2$ of the lag model from $0.66$ to $0.28$ at URL grain and from $0.55$ to $0.19$ at domain grain.
\item \emph{The longest lags had one pair each,} both anchored on the first day (Figure~\ref{fig:ladder}). Where the between-pair variation could be measured (lags 1 to 6), the standard deviation of pair means reached $0.086$ at lag 3, larger than the $\pm 0.038$ half-width of the query-only interval. The retention at lag 8 is one realization of one day pair and is better reported as a bracket of roughly $[0.44, 0.56]$.
\item \emph{The asymptote was not identified.} A model $a + b\,e^{-\kappa\Delta}$ fitted the eight lags about as well as the log-linear model (residual sum of squares $0.0033$ at URL grain) with $a$ at the boundary of its range ($0.06$).
\item \emph{The normalization was described as conservative.} The preregistration stated that dividing by the noise floor would bias the projection toward the null. Under its free-intercept fit this is false (Proposition~\ref{prop:intercept}).
\end{enumerate}
The non-identified projections appear here once, so that the failure is visible: the design yielded raw-retention half-lives of 30.2 days at URL grain (interval $[21.7, 47.7]$) and 51.2 days at domain grain ($[31.8, 123.0]$), and a projected 28-day URL retention of $0.309$ ($[0.243, 0.384]$). For the reasons above these are not findings. The intervals reflect query resampling under a functional form known to be misspecified, and the bias of that form runs downward (Section~\ref{sec:simulation}).

\paragraph{What the design did identify.} Two quantities need no model and no extrapolation (Table~\ref{tab:core}). The stable core is substantial at every threshold, and its size depends on the threshold strongly enough that a single value would mislead. The recurrence response, presence on the last day given the number of the first three days on which a source appeared, is steep and monotone at both grains. Among sources that recurred on all three early days, URL and domain grain differ by about $0.02$, so the larger whole-set difference between grains comes largely from the composition of the single-appearance fringe. The folk claim that ``URLs churn but domains persist'' is not what these data show; at both grains, the fringe churns and the recurring core persists.

\begin{table}[t]
\centering
\small
\caption{Quantities the preregistered panel identifies without a model. Top: the stable core as a threshold family, mean core size per query and its coverage of a day's cited domain set (94 queries with at least five observed days). Bottom: recurrence response, presence on the last collection day by the number of the first three days on which a source appeared, with Wilson 95\% intervals that ignore clustering by query and are therefore too narrow.}
\label{tab:core}
\begin{tabular}{@{}lrr@{}}
\toprule
Threshold (days of 6) & Core size & Coverage \\
\midrule
$\geq 3$ & 11.00 & 0.693 \\
$\geq 4$ & 7.86 & 0.569 \\
$\geq 5$ & 5.82 & 0.460 \\
$\geq 6$ & 3.78 & 0.329 \\
\bottomrule
\end{tabular}\\[8pt]
\begin{tabular}{@{}lrlrl@{}}
\toprule
Early days & \multicolumn{2}{c}{Domains} & \multicolumn{2}{c}{URLs} \\
 & $n$ & last day & $n$ & last day \\
\midrule
1 of 3 & 1{,}181 & 0.181 [0.160, 0.204] & 1{,}473 & 0.151 [0.133, 0.170] \\
2 of 3 & 394 & 0.553 [0.504, 0.602] & 449 & 0.523 [0.477, 0.569] \\
3 of 3 & 464 & 0.843 [0.807, 0.873] & 445 & 0.820 [0.782, 0.853] \\
\bottomrule
\end{tabular}
\end{table}

\paragraph{A disclosure about provenance.} The preregistration was committed to version control on its own, five minutes before the analysis commit and as its ancestor, and was not edited afterwards. The decision rule kept branches that could have returned an undecided outcome, and one of them came within $0.064$ of firing. The panel's row in our internal hypothesis registry, however, was added retrospectively, after the result was known, to close a gap in that registry. A commit graph shows that a preregistration existed before an analysis was committed; it cannot show that no analysis was run before the preregistration was written. We report both facts so that a reader can weigh them.

\section{A reporting checklist for citation and AI visibility numbers}
\label{sec:checklist}

A citation-stability or citation-rate figure cannot be interpreted unless it declares the following.\footnote{Free online calculators from the authors for three related computations: how many queries and repeated draws a measurement needs (\plainlink{https://www.elementera.com/tools/ai-visibility-sample-size?r=iTZzRC}{AI visibility sample size}), whether a change between two measurements exceeds draw noise (\plainlink{https://www.elementera.com/tools/ai-visibility-change-test?r=iTZzRC}{AI visibility change test}), and share of voice with a margin of error (\plainlink{https://www.elementera.com/tools/ai-share-of-voice-calculator?r=iTZzRC}{AI share of voice calculator}).} The list follows from Sections~\ref{sec:setting} to~\ref{sec:observability}.

\begin{enumerate}
\item \textbf{Grain}: URL, host or registrable domain, and the canonicalization rules.
\item \textbf{Metric}: retention (with the anchor) or Jaccard or another overlap, and whether rank is used.
\item \textbf{Union convention and $k$}: single draw or union of $k$ draws per occasion; if $k=1$, a statement that the noise floor is not identified by the design.
\item \textbf{Occasions and span}: dates of the first and last occasion, the occasion layout, and the number of occasion pairs at each reported lag that do not share an anchor.
\item \textbf{Collapse rule}: how several occasions become one label, if they do.
\item \textbf{Inference}: whether intervals resample occasions as well as queries.
\item \textbf{Missing outcomes}: how failed captures and answers without sources were scored, and any change in that rule during the panel.
\item \textbf{Query population}: how queries were chosen and weighted, and whether results are per query or pooled.
\item \textbf{Capture path}: engine, interface or API, model identity as reported by the response, locale, device, data provider, and what the instrument records.
\item \textbf{Conditioning}: whether a reported rate conditions on appearing in a rendered list, and whether the funnel stages are nested on the fields used.
\item \textbf{Extrapolation}: whether any reported horizon exceeds the observed span.
\end{enumerate}

\section{Related work}
\label{sec:related}

\paragraph{Optimization for generative engines.} \citet{aggarwal2024geo} introduced generative engine optimization and measured how rewriting a source changes its share of a generated answer when the source is placed in a fixed context of retrieved documents. Later work tested such methods across domains and with several competing actors \citep{puerto2025cseo} and in large controlled comparisons of two injected sources \citep{vishwakarma2026whatgetscited}. These designs fix admission by construction and identify effects at the second stage of the funnel for the apparatus used. A recent survey describes the pipeline as stochastic and partially observable, writes the citation probability as a product of activation, retrieval and citation given retrieval, and recommends defining the estimand before data collection and clustering inference by query, source and date \citep{martinez2026survey}. We agree with each of these recommendations. Our contribution is to state the nesting axiom behind the product, to examine whether the exposed fields on several surfaces can serve as the retrieval term, to argue that the rendered list is a collider, and to give identification results for repeated-query designs.

\paragraph{Measurement of live answer engines.} \citet{schulte2026dontmeasureonce} measured day-to-day overlap of cited sources across four engines and compared it with same-day re-runs, concluding that one-off observations are unreliable and that visibility should be treated as a distribution. In our terms their comparison is a design with $k \ge 2$, and Proposition~\ref{prop:k} explains why the comparison is necessary. \citet{grossman2026disrupts} compared Google search, AI Overviews and Gemini on a public benchmark of 11{,}500 queries, found that the sources differ substantially between surfaces, and found AI Overviews less consistent across two runs of the same query. \citet{xu2026aio} measured AI Overview activation, source quality and claim fidelity over a 40-day window. Earlier work audited whether the citations of generative search engines support the statements they are attached to \citep{liu2023verifiability}, and \citet{gienapp2024generative} discussed how evaluation of generative retrieval must change when the output is text rather than a ranking.

\paragraph{Reliability of retrieval evaluation.} Concern with whether evaluation numbers are stable under changes in the measurement process is old in information retrieval, from the reliability of pooled test collections \citep{zobel1998reliable} to the sensitivity of system comparisons to relevance assessors \citep{voorhees2000variations}. Answer engines add a source of variation that test collections do not have: the system itself is stochastic between requests.

\paragraph{Measurement models and estimands.} \citet{jacobs2021measurement} argued that computational systems operationalize unobservable constructs through measurement models whose assumptions must be made explicit, and \citet{lundberg2021estimand} argued that a study should define its target quantity before choosing an estimator. Imperfect detection and repeat visits are standard in ecology \citep{mackenzie2002occupancy,mackenzie2003dynamic,royle2007dynamic}, and our latent-membership model is an instance of that framework. Conditioning on a common effect is a well-studied source of bias \citep{berkson1946,greenland2003collider,hernan2004selection,cole2010collider,elwert2014endogenous,pearl2009causality}, and inference for crossed designs is treated by \citet{owen2007pigeonhole} and \citet{cameron2011multiway}. Preregistration is discussed by \citet{nosek2018preregistration}.

\section{Limitations and scope}
\label{sec:limitations}

\paragraph{Action effects are outside the framework.} Every estimand in this paper describes the observation process. None identifies the effect of an action, and the reason is structural.\footnote{Tests of popular AEO beliefs run by the authors are summarized on the \plainlink{https://www.elementera.com/claims-we-tested?r=iTZzRC}{AEO claims we tested} page.}

\begin{proposition}[Records without actions]
\label{prop:actions}
Let a record contain, for each unit, a state $S$, a prediction or recommendation $P$ and an outcome $O$, but not the action $D$ actually taken. Then the average effect $\E[O(1) - O(0)]$ is not identified from the distribution of $(S, P, O)$.
\end{proposition}

\begin{proof}
Take any distribution of $(S,P,O)$. Construct two data-generating processes that both reproduce it: in the first, $D$ is independent of everything and $O(1) = O(0) = O$; in the second, $D$ is set to 1 on a subset of units of positive probability chosen independently of $(S,P,O)$, with $O(1) = O$ and $O(0) = O - \delta$ for some $\delta \neq 0$ on those units, and $O(1) = O(0) = O$ elsewhere. The observed outcome is $O$ in both processes and $D$ is not recorded, so both induce the same distribution of $(S,P,O)$, yet the average effect is $0$ in the first and $\delta\,\Prob(D=1) \neq 0$ in the second.
\end{proof}

The point is not one of power. We checked it in simulation on panels matched to the structure of one of our own (37{,}808 simulated transitions): an estimator that does not see which cells were acted on had power $0.000$ to detect injected effects of 5, 10 and 20 percentage points in every condition tested (500 simulations per condition), while an estimator given the true assignment detected the same injected effects with power rising from $0.56$ at 5 points to $1.00$ at 20 points. Identifying an action effect requires a verified record of the action taken, and even then voluntary action is not assignment; a randomized design is outside the scope of this paper.

\paragraph{Other limits.}
\begin{itemize}
\item \emph{Engines, locales and dates.} Our exhibits come from captures made in July 2026, in US English, through specific APIs and data providers. Engines change. No cross-engine claim in this paper rests on more than the cited studies, and each exhibit is scoped to the surface on which it was measured.
\item \emph{Sample sizes.} The render-dependence test of Section~\ref{sec:observability} is a pilot with 18 query cells and cannot support a verdict. The preregistered panel of Section~\ref{sec:panel} has 100 queries and six collection days and was not designed for per-vertical analysis; we report none.
\item \emph{General identifiability.} We prove identification and non-identification for specific model classes. We have not characterized, for general latent dynamics, the budgets $(k, d, n)$ under which window length and per-draw inclusion are separately identified. The simulation study is evidence of recovery under one model, not a proof.
\item \emph{The collider argument.} The structure in Figure~\ref{fig:dag} is a hypothesis about how engines produce answers. We have not adjudicated it, and the evidence on the public surfaces is consistent with it without establishing it.
\item \emph{External audit.} The analyses have been reviewed internally only. We invite replication of the designs and of the checklist.
\item \emph{Data release.} The underlying captures are not released with this version (see the reproducibility statement).
\end{itemize}

\section{Conclusion}
\label{sec:conclusion}

A citation number from an AI answer engine, whether a citation rate, a stability figure or an AI search visibility score used for AEO or GEO, is a statement about one query, capture path, occasion and draw that has been averaged over draws, occasions, queries, capture paths and the engine's own expansion of the query. Some of these averages can be undone by design. Repeating each query at least twice per occasion identifies the noise floor. Long spans with many unanchored pairs identify how fast citations decay. Resampling occasions as well as queries gives intervals that mean what they say. One average cannot be undone from outside the engine: its expansion of a user query into sub-queries. On the surfaces we have captured, the fields that look like a retrieval pool are attribution lists, so the admission stage of the citation funnel is not identified from what the engines expose, and estimates that condition on the rendered list should say so. We hope the reporting checklist in Section~\ref{sec:checklist} makes these choices visible in future reports, including ours.

\section*{Reproducibility statement}

The simulation study (Section~\ref{sec:simulation}, Figure~\ref{fig:sim}) is produced by \texttt{anc/simulation.py}, released with this preprint as an ancillary file. The script is deterministic, requires Python with NumPy and Matplotlib, makes no network calls, and writes every simulation number quoted in Section~\ref{sec:simulation} to \texttt{simulation\_summary.json}. Figure~\ref{fig:ladder} is produced by a companion script in the authors' repository, which also records, with source paths, the numbers from our data in Tables~\ref{tab:surfaces} and~\ref{tab:core}, the sub-query counts of Section~\ref{sec:ladder}, the pilot of Section~\ref{sec:observability} and the collapse-rule counts of Section~\ref{sec:model}; the remaining numbers from our data are taken from the committed analysis records listed in Appendix~\ref{app:provenance}. Figure~\ref{fig:ladder} and Table~\ref{tab:core} are built from analysis outputs of the preregistered panel. Within the authors' repository, that panel has a provenance chain consisting of the preregistration, checksummed raw responses, and a recompute recipe pinned to a code commit; the other exhibits are derived from checksummed captures by scripts in the same repository. The raw captures are not released with this version, because their redistribution depends on the terms of the data providers and on our internal clearance process.

\section*{Conflict of interest}

All four authors are founders of \plainlink{https://www.elementera.com/?r=iTZzRC}{Elementera AI Inc.}, whose products are \plainlink{https://www.elementera.com/product/lenz?r=iTZzRC}{Lenz}, which measures the visibility of brands and sources in AI answer engines, and \plainlink{https://www.elementera.com/product/originum?r=iTZzRC}{Originum}, which turns what experts know into content and gives each draft an originality score; the company also sells \plainlink{https://www.elementera.com/managed-aeo-services?r=iTZzRC}{managed services} built on both. The founders' earlier peer-reviewed research is listed on the \plainlink{https://www.elementera.com/research-papers?r=iTZzRC}{research papers} page. The methods discussed here bear directly on how such measurements, including those sold by Elementera, should be designed and reported. One exhibit (Section~\ref{sec:panel}) is a preregistered test that failed against a specification of our own product. This work was funded by Elementera. Search and answer-engine captures were purchased by Elementera through commercial data-provider and API accounts; part of the model computation used Microsoft Azure sponsorship credits granted to Elementera. No external grant supported this work.

\bibliographystyle{plainnat}
\bibliography{refs}

\appendix
\section{Provenance of exhibits}
\label{app:provenance}

Internal identifiers refer to the authors' hypothesis registry or, where no registry row exists, to the analysis artifact. ``Internal review not complete'' means that the internal review of a finding, a step after its record is complete, has not yet been done; such results are reported as descriptive.

\begin{center}
\small
\begin{tabular}{@{}p{4.6cm}p{3.6cm}p{5.4cm}@{}}
\toprule
Exhibit & Internal identifier & Status \\
\midrule
Simulation study (Section~\ref{sec:simulation}, Figure~\ref{fig:sim}) & none (synthetic) & Synthetic; code released \\
Observed vs.\ latent rotation (Section~\ref{sec:model}) & H-0157, H-0158, H-0161 & Internal review of the emission model not complete \\
Universe and admission share (Section~\ref{sec:model}) & H-0153 & Internal review not complete \\
Collapse rules on captured data (Section~\ref{sec:model}) & render-proxy calibration, 2026-08-07 & No registry row \\
Within- and across-occasion agreement (Section~\ref{sec:identification}) & H-0094 & Record complete \\
Exposed sub-queries (Section~\ref{sec:ladder}) & computed by \texttt{make\_figures.py} & No registry row \\
Worked example quotation (Section~\ref{sec:ladder}) & H-0254 & Quoted verbatim from our public reporting \\
Exposed source fields (Table~\ref{tab:surfaces}) & engine degeneracy sweep; RENDER-DEGENERACY-01 & No registry row \\
Render-dependence pilot (Section~\ref{sec:observability}) & render-collider v2 & No registry row \\
Preregistered panel (Section~\ref{sec:panel}) & H-0209 & Record complete \\
Action-effect power check (Section~\ref{sec:limitations}) & H-0184 & Record complete \\
\bottomrule
\end{tabular}
\end{center}

\end{document}